\documentclass[%
 reprint,
 superscriptaddress,
 showpacs,
 preprintnumbers,
 amsmath,amssymb,
 aps,
 prd,
 floats,
 nofootinbib,
]{revtex4-1}

\usepackage{dcolumn}
\usepackage{bm}

\usepackage{amsfonts}
\usepackage{latexsym}

\usepackage{float}

\def\bequ{\begin{equation}}
\def\eequ{\end{equation}}
\def\barr{\begin{array}}
\def\earr{\end{array}}

\def\ben{\begin{equation}}
\def\een{\end{equation}}
\def\bena{\begin{eqnarray}}
\def\eena{\end{eqnarray}}

\def\b1{e^0}

\newcommand{\be}{\begin{equation}}
\newcommand{\ee}{\end{equation}}
\def\bea{\begin{eqnarray}}
\def\eea{\end{eqnarray}}

\def\del {\partial}

\def\be{\begin{equation}}
\def\ee{\end{equation}}
\def\bea{\begin{eqnarray}}
\def\eea{\end{eqnarray}}

\def\lesssim{\mathrel{\hbox{\rlap{\hbox{\lower4pt\hbox{$\sim$}}}\hbox{$<$}}}}
\def\gtrsim{\mathrel{\hbox{\rlap{\hbox{\lower4pt\hbox{$\sim$}}}\hbox{$>$}}}}

\begin{document}

\preprint{UPR-1262-T}

\title{Addentum to: Conformal Invariance and  Near-extreme Rotating AdS Black Holes}

\author{Tolga Birkandan}
 \affiliation{Department of Physics and Astronomy, University of Pennsylvania, Philadelphia, PA 19104, USA.}
 \affiliation{Department of Physics, Istanbul Technical University, Istanbul 34469, Turkey.}

\author{Mirjam  Cveti\v c}
 \affiliation{Department of Physics and Astronomy, University of Pennsylvania, Philadelphia, PA 19104, USA.}
 \affiliation{Department of Physics, Istanbul Technical University, Istanbul 34469, Turkey.}
 \affiliation{Center for Applied Mathematics and Theoretical Physics, University of Maribor, Maribor, Slovenia}

\begin{abstract}
We  obtained retarded Green's functions for massless scalar fields in the background of near-extreme, near-horizon rotating charged black hole of five-dimensional minimal gauged supergravity in Phys.\ Rev.\ D {\bf 84}, 044018 (2011).  For general nonextreme black holes,  we also derived the radial part of the  massless Klein-Gordon equation  with zero azimuthal-angle eigenvalues, and  showed that it  is a general Heun's equation with a regular singularity at each horizon $u_k$ ($k=1,2,3$) and  at  infinity. We  derived explicitly that the residuum of a pole at each $u_k$ is associated with  the surface gravity there. In this addendum, probing regular singularities  at each $u_k$  we  complete the derivation  of the  full radial equation with nonzero azimuthal-angle eigenvalues. The residua now include modifications by  the angular velocities at respective horizons. This result completes the  analysis of the  wave equation for  the massless Klein-Gordon equation for the general rotating charged black hole of five-dimensional minimal gauged supergravity.
\end{abstract}


\maketitle

In \cite{Birkandan:2011fr}, we studied the massless Klein-Gordon equation  for the general nonextreme rotating black hole in minimal five-dimensional gauged supergravity, given by Chong et al. in \cite{Chong05}.  The black hole is specified by the mass, two angular momenta, three equal charges and a cosmological constant, parametrized by  $m,\ a,\ b, \  q$, and $g$.\footnote{The reader should  refer to  \cite{Birkandan:2011fr} for the  explicit form of the metric  and its parametrization.}
We showed that the wave equation was separable, and obtained the general polar angle equation.  As for the radial part,  we obtained the explicit equation  only for  zero  eigenvalues ($m_1=m_2=0$) of the azimuthal angle coordinates   $\phi $ and $\psi $  in the scalar field Ansatz
\begin{equation}
\Phi =e^{-i\omega t+im_{1}\phi +im_{2}\psi }R(u)F(y).
\end{equation}
It was shown in the original paper that for $m_1=m_2=0$  the radial part of the Klein-Gordon equation could be cast in the form
\begin{equation}
\begin{split}
\frac{d}{du}\left( \Delta _{u}\frac{dR}{du}\right)
+\frac{1}{4}\bigg\{ \bigg[
\frac{n_{1}}{\kappa _{1}^{2}(u-u_{1})}+\frac{n_{2}}{\kappa _{2}^{2}(u-u_{2})}%
\\+\frac{n_{3}}{\kappa _{3}^{2}(u-u_{3})}+Gn_{4}\bigg] \omega
^{2}-c_{0}\bigg\} R=0,\label{radial}\,
\end{split}
\end{equation}
%
where $\kappa_{k}$ ($k=1,2,3$) are the surface gravities associated with the  three horizons $u_{k}$ ($k=1,2,3$) , which are the solutions of the horizon equation that can be written in the form: $\Delta_{u}=G(u-u_{1})(u-u_{2})(u-u_{3})$. It should be remembered that we have changed the radial coordinate as $u=r^2$. $G \equiv g^2$ is related to the cosmological constant $\Lambda=-6G$, $c_0$ is the separation constant and $n_k$ are the constants given in the original article as
\begin{eqnarray}
n_{1} &=&G(u_{1}-u_{2})(u_{1}-u_{3}), \\
n_{2} &=&-G(u_{1}-u_{2})(u_{2}-u_{3}), \\
n_{3} &=&G(u_{1}-u_{3})(u_{2}-u_{3}), \\
n_{4} &=&\frac{1}{G^{2}}.
\end{eqnarray}
It turned out to be  difficult to work with the full radial equation with nonzero azimuthal-angle  eigenvalues ($\{m_1,\ m_2\}\ne 0$), in particular to extract the explicit structure of the residuum  of a  pole at each $u_k$  ($k=1,2,3$) in terms of the surface gravity $\kappa_k$  and angular velocity $\Omega_k$ which are of the form
\begin{eqnarray}
\kappa _{1} &=&\frac{G(u_{1}-u_{3})(u_{1}-u_{2})\sqrt{u_{1}}}{%
(u_{1}+a^{2})(u_{1}+b^{2})+abq}, \label{surface1} \\
\kappa _{2} &=&\frac{G(u_{2}-u_{3})(u_{1}-u_{2})\sqrt{u_{2}}}{%
(u_{2}+a^{2})(u_{2}+b^{2})+abq}, \label{surface2} \\
\kappa _{3} &=&\frac{G(u_{2}-u_{3})(u_{1}-u_{3})\sqrt{u_{3}}}{%
(u_{3}+a^{2})(u_{3}+b^{2})+abq},\label{surface3}
\end{eqnarray}
and
\begin{eqnarray}
\Omega_{ak}&=&\frac{(1-a^{2}G) [a(u_k+b^2)+bq]}{(u_k+a^2)(u_k+b^2)+abq},\label{veloc1} \\
\Omega_{bk}&=&\frac{(1-b^{2}G) [b(u_k+a^2)+aq]}{(u_k+a^2)(u_k+b^2)+abq}.\label{veloc2}
\end{eqnarray}
In this addendum, we will seek a form of the general radial equation, when  $\{m_1,m_2\}\ne 0$, as
\begin{equation}
\frac{d}{du}\left( \Delta _{u}\frac{dR}{du}\right) +\left( \frac{n_{1}\alpha
_{1}^{2}}{u-u_{1}}+\frac{n_{2}\alpha _{2}^{2}}{u-u_{2}}+\frac{n_{3}\alpha
_{3}^{2}}{u-u_{3}}+\alpha _{4}\right) R=0,
\end{equation}
where $\alpha _{k}$ ($k=1,2,3$) will  now include the surface gravities and the angular velocities for the associated horizons.

Let us  probe the radial equation for the regular singular points. We need to consider the radial equation as an equation in the complex $r$-plane. Then the solutions will have branch cuts at the regular singular points which can be seen by performing a series solution around these points. The solutions will describe outgoing and ingoing waves at the associated horizon. For a general equation in the form
\begin{equation}
[\del_r \Delta(r) \del_r - V(r)]R(r)=0,
\end{equation}
we have
\begin{eqnarray}
R_{r_{*}}^{out} &=&(r-r_{*})^{i\alpha _{*}}[1+\mathcal{O}(r-r_{*})], \\
R_{r_{*}}^{in} &=&(r-r_{*})^{-i\alpha _{*}}[1+\mathcal{O}(r-r_{*})],
\end{eqnarray}
for the regular singularity at $r=r_{*}$ other than infinity. It turns out that the exponent in this expansion has a special form, e.g.,  for the Kerr case one obtains
\begin{equation}
\alpha_{*}=\frac{\omega-\Omega_{*}m}{2\kappa_{*}},
\end{equation}
which is needed for the formal wave equation analysis \footnote{This result also applies \cite{Cvetic:1997xv,Cvetic:1997uw} to general rotating multicharged black holes in maximally supersymmetric ungauged supergravities \cite{Cvetic:1996kv,Cvetic:1996xz}.}. This observation is also key  to our study of $\alpha_{*}$  as it is also the exponent that is used to form the monodromy matrix associated with that singularity. The details and applications of this method can be found in \cite{Castro:2013lba,Castro:2013kea}.

We need only to probe for the horizon singularities, namely $u_{1},u_{2}$ and $u_{3}$ for our analysis.

For the regular singular points $u_{k}$ we will have,
\begin{eqnarray}
R_{u_{k}}^{out} &=&(u-u_{k})^{i\alpha _{k}}[1+\mathcal{O}(u-u_{k})], \\
R_{u_{k}}^{in} &=&(u-u_{k})^{-i\alpha _{k}}[1+\mathcal{O}(u-u_{k})].
\end{eqnarray}
Using this expansion in the radial equation, and after some algebra we find $\alpha _{k}$ as
\begin{equation}
\alpha _{k}=\pm \left( \frac{\omega }{2\kappa _{k}}-\frac{\Omega _{ak}}{%
2\kappa _{k}}m_{1}-\frac{\Omega _{bk}}{2\kappa _{k}}m_{2}\right).
\end{equation}
Here, $\kappa _{k}$'s are the surface gravities  (\ref{surface1}-\ref{surface3}) and $\Omega _{ak}$ and $\Omega _{bk}$ are the two  angular velocities  (\ref{veloc1},\ref{veloc2}) \cite{Chong05,Birkandan:2011fr}. Probing for a single regular singularity makes it  possible to determine the structure due to the associated angular velocities, once one has the surface gravities.

Using the original radial equation we can read off  the $n_{k}$ constants and the  $\alpha _{4}$ term in the full equation as
\begin{eqnarray}
n_{1} &=&G(u_{1}-u_{2})(u_{1}-u_{3}), \\
n_{2} &=&-G(u_{1}-u_{2})(u_{2}-u_{3}), \\
n_{3} &=&G(u_{1}-u_{3})(u_{2}-u_{3}),
\end{eqnarray}
and
\begin{equation}
\alpha _{4}=\frac{1}{4}(\frac{\omega ^{2}}{G}-c_{0}),
\end{equation}
which are the same in the equation without the angular velocities.

In conclusion, we have obtained the full radial equation with explicit contributions from  surface gravities and angular velocities by probing the regular singular points (associated with the three horizons of the black hole) for the monodromy exponents with the method, briefly described above. This result completes the  analysis of the full wave equation, initiated in \cite{Birkandan:2011fr}.

\begin{acknowledgements}

T.B. would like to thank the Department of Physics and Astronomy at the University of Pennsylvania for hospitality during the early course of this work. The research of T.B. is supported by TUBITAK, the Scientific and Technological Council of Turkey, Istanbul Technical University [\.{I}T\"{U}\ BAP 37519] and High Energy Theory Fund of the Department of Physics and Astronomy at the University of Pennsylvania. M.C. would like to thank the Department of Physics at the Istanbul Technical University for hospitality during the course of this work. M.C.'s visit to Istanbul Technical University was supported by TUBITAK ``2221-Fellowship for Visiting Scientists" with number 1059B211400897. M.C. is also supported by the DOE Grant No. DOE-EY-76-02- 3071, the Fay R. and Eugene L. Langberg Endowed Chair, the Slovenian Research Agency (ARRS), and the Simons Foundation Fellowship.
\end{acknowledgements}

\end{document}